\documentclass[useAMS,usenatbib,preprintnumbers,footinbib]{revtex4}
\usepackage{epsfig}
\input psfig.sty
\usepackage{graphicx}
\usepackage{dcolumn}
\usepackage{bm}
\usepackage{amsmath} 
\usepackage[english]{babel}
\usepackage{mathrsfs}
\usepackage{amsfonts}
\usepackage{amssymb}
\usepackage[latin1]{inputenc}
\usepackage{hyperref}
\usepackage{longtable}
\usepackage{float}
\usepackage{dcolumn}
\usepackage{appendix}
\usepackage{multirow}
\usepackage{color}

\newcommand{\mytilde}{\raise.17ex\hbox{$\scriptstyle\mathtt{\sim}$}}
\newcommand{\barr}{\begin{eqnarray}}
\newcommand{\earr}{\end{eqnarray}}
\newcommand{\bea}{\begin{eqnarray*}}
\newcommand{\eea}{\end{eqnarray*}}
\newcommand{\beq}{\begin{equation}}
\newcommand{\eeq}{\end{equation}}
\newcommand{\tc}{\eta_{\vec{k}}^c}
\newcommand{\nn}{\nonumber \\}
\newcommand{\zk}{|z_k|}
\newcommand{\mH}{\mathcal{H}}

\newcommand{\dphi}{\delta \phi}
\newcommand{\x}{\vec{x}}

\newcommand{\bra}{\langle}
\newcommand{\ket}{\rangle}
\newcommand{\nk}{\vec{k}}
\newcommand{\ann}{\hat{a}} 
\newcommand{\cre}{\hat{a}^{\dagger}}    

\newcommand{\mR}{\mathcal{R}}

\newcommand{\Mpc}{\mathrm{Mpc}}

\newcommand{\A}{$\mathcal{A}$~}
\newcommand{\B}{$\mathcal{B}$~}

\begin{document}

\title{Constraining quantum collapse inflationary models with CMB data}

\author{Micol Benetti$^1$\footnote{E-mail: micolbenetti@on.br}}

\author{Susana J. Landau$^2$\footnote{E-mail: slandau@df.uba.ar }}

\author{Jailson S. Alcaniz$^1$\footnote{E-mail: alcaniz@on.br}}

\affiliation{$^1$Departamento de Astronomia, Observat\'orio Nacional, 20921-400, Rio de Janeiro - RJ, Brasil}

\affiliation{$^2$Departamento de F\'isica, Facultad de Ciencias Exactas y Naturales, Universidad de Buenos Aires and IFIBA, CONICET, Ciudad Universitaria - PabI, Buenos Aires 1428, Argentina}

\date{\today}

\begin{abstract}
The hypothesis of the self-induced collapse of the inflaton wave function was proposed as responsible for the emergence of inhomogeneity and anisotropy at all scales. 
This proposal was studied within an almost de Sitter space-time approximation for the background, which  led to a perfect scale-invariant power spectrum, and also  for a quasi-de Sitter background,  which allows to distinguish departures from the standard approach due to the inclusion of the collapse hypothesis.
In this work we perform a Bayesian model comparison for two different choices of the self-induced collapse in a full quasi-de Sitter expansion scenario.
In particular, we analyze the possibility of detecting the imprint of these collapse schemes at low multipoles of the anisotropy temperature power spectrum of the Cosmic Microwave Background (CMB) using the most recent data provided by the Planck Collaboration. Our results show that one of the two collapse schemes analyzed provides the same Bayesian evidence of the minimal standard cosmological model $\Lambda$CDM, while the other scenario is weakly disfavoured with respect to the standard cosmology.
\end{abstract}
\keywords{Cosmic Microwave Background Radiation theory, Cosmological parameters from CMBR, Physics of the early universe, Inflation}

\maketitle
\section{Introduction}
According to the standard inflationary paradigm, the origin of the cosmic structures is explained  by a background Friedmann-Robertson-Walker (FRW) cosmology with a nearly exponential expansion driven by the potential of a single scalar field and from its quantum fluctuations characterised by a simple vacuum state. 
However, when this picture is considered more carefully, a conceptual issue arises: while the initial state  that characterises the quantum perturbations of both the inflaton field and the metric is highly homogeneous and isotropic, the present state of the universe is described by a state  with  inhomogeneities and anisotropies. 
As known, the quantum unitary evolution  can not be responsible for breaking the initial symmetries of the early quantum state. This issue has been discussed in previous papers~\cite{PSS06,Shortcomings,US08,Leon10,DT11,Leon11,LSS12,LLS13,LLP15,CPS13}, where the {\it collapse proposal}  has been developed. 
The key ingredient of this proposal is to assume a self-induced collapse of the inflaton wave function as the responsible for the emergence of inhomogeneities and anistropies at each particular length scale.

The idea that the collapse of the wave function could be regarded as an actual physical process induced by gravity  was proposed in Refs. 
\cite{penrose1996,diosi1987,diosi1989,ghirardi1985}. 
On the other hand, various proposals of that sort have been developed for studying problems in different context than the cosmological one \cite{ghirardi1985,pearle1989,bassi2003,bassi2013}. 
These proposals  might well be compatible with the self-induced collapse of the inflaton's wave function that we are considering. Here, the  hypothesis simply  assumes that something intrinsic to the system, i.e., independent of external 
agents (e.g., observers), triggers the collapse or reduction of the quantum  mechanical state of the system. 
The proposal is, at this point, a purely phenomenological scheme. 
It does not attempt to describe the process in terms of some specific new physical theory, but just to provide a general parameterisation of the quantum transition involved.  It is worth mentioning that the previous conceptual problem is sometimes known in the literature as the quantum-to-classical transition of the primordial perturbations. In this context, some authors have argued that decoherence \cite{kiefer,grishchuk} can give a good explanation of the emergence of anisotropies and inhomogeneities.  Other approaches  seem to adopt the Everett  ``many-worlds'' 
interpretation of quantum mechanics when confronted with this problem. However, it has been shown that neither decoherence \cite{adler,schlosshauer} nor the Everettian formulation can solve the quantum measurement problem (we refer the reader to \cite{PSS06,Shortcomings,Leon10,Leon11,LLS13} for a detailed discussion of this issue). 

In order to treat the collapse process, we assume that at a certain stage in cosmic evolution there is an induced jump in a state describing a particular mode of the quantum field, in a such similar  way of a  quantum mechanical collapse of the wave function associated with a measurement. However in our scheme there is no external measuring device or observer responsible for triggering such collapse. The next issue is to define the characteristic of the state in which such jump occurs. In particular, we need a criterion  to determine the expectation values of the field and the momentum conjugate variables for the post-collapse state, without relying on some particular collapse mechanism. In previous works, \cite{PSS06,US08,Leon11,LLP15} various possibilities regarding the description of the quantum expectation values in the post-collapse state were developed. 
We will refer to them as {\it collapse schemes} and we focus in this work ones called {\it Newtonian} and {\it Wigner}. 

In a previous work~\cite{LLP15}, some of us have calculated the primordial power spectrum for  different collapse schemes in a full quasi-de Sitter background, and obtained an expression of the form $P(k) = A_s k^{n_s - 1 } Q(k)$  where  $Q(k)$ is a 
function introduced by the collapse hypothesis. Furthermore, it has been shown \cite{US08,LSS12,LLP15} that the primordial power spectrum is similar to the one predicted by the standard inflationary model if the conformal time of collapse of each mode of the field is given by $\tc = {\mathcal{A}}/{k}$ with \A being a constant. In other words, in such case both the  standard power-law prediction of the primordial power spectrum and the angular power spectrum of the CMB temperature and polarization are recovered. Departures from this expression were also studied~\cite{LLP15}, e.g., $\tc = {\mathcal{A}}/{k}+ {\rm const}$. For this case, it was shown that the primordial power spectrum is significantly modified  with respect to the standard prediction for values of $k>10^{-3}$. Furthermore, it was also studied the effect of the collapse hypothesis on the CMB temperature power spectrum, which showed an increment in the secondary peaks for increasing values of the constant. 

In this work, we discuss the observational viability of the {\it Newtonian} and {\it Wigner} collapse scheme scenarios in the light of the Planck 2015 data. We work with a new parameterisation of the collapse time, $\tc = {\mathcal{A}}/{k}+{\mathcal{B}}/{k^2}$, which, differently from the previous expressions, is able to produce modifications over the entire multipole interval of the primordial power spectrum. In particular, significant departures from the standard prediction are observed at low-$\ell$, providing a possible explanation for  the lack of power in the CMB temperature anisotropies at large angular scales, as recently confirmed by the Planck data~\cite{Ade:2015xua, Copi:2013cya}. We perform a
Bayesian analysis using both the Metropolis-Hastings algorithm implemented in {\sc CosmoMC} and the nested sampling algorithm of {\sc Multinest}.  We find that  the {\it Wigner} collapse scheme scenario provides the same Bayesian evidence of the minimal standard cosmological model ($\Lambda$CDM) \footnote{By standard cosmological model ($\Lambda$CDM) we understand a specific choice of the cosmological parameters plus the standard inflationary model as opposite to the collapse models, where the collapse hypothesis is assumed for inflation and the cosmological parameters remain unchanged}, while the {\it Newtonian} case is weakly disfavoured with respect to the standard cosmology.

The paper is organized as follows: In Sec. \ref{model}, we briefly review the collapse hypothesis within the semiclassical gravity approximation and summarize the procedure to obtain the post-collapse states; in Sec. \ref{primordialpower} we review the expressions for the primordial power spectrum calculated in Ref. \cite{LLP15} and discuss the effect of  the new parameterisation for the collapse time on the primordial power spectrum for the {\it Newtonian} and {\it Wigner} schemes. Furthermore, we also analyse the effect of the proposed parametrisation on the CMB temperature angular spectrum. In Sec. \ref{Method} we introduce the computational and statistical tools and the data set used in our analysis. 
In Sec. \ref{results} we present the results of our analysis and the constraints on the cosmological and collapse parameters. Finally, in Sec. \ref{conclusions}, we summarize the main results of the paper and present our conclusions.
\section{The model}
\label{model}

In this section, we briefly review the key aspects of inflationary models with a self-induced colapse of the inflaton's wave function. In particular, we focus on the models analyzed in Ref. \cite{LLP15}, where no particular collapse mechanism was assumed. Regarding notation and conventions, we will work with signature  $(-,+,+,+)$ for 
the metric; primes over functions will denote derivatives with respect to the 
conformal time $\eta$, and we will use units  where $c=\hbar=1$ but keep the 
gravitational constant $G$.  As in standard inflationary models, we focus on the action of a single scalar field, minimally coupled to gravity, with an appropriate potential: 
\begin{eqnarray}
S[\phi,g_{ab}] = \int d^4x \sqrt{-g} \bigg[ \frac{1}{16 \pi G} R[g] 
-\frac{1}{2}\nabla_a \phi \nabla_b \phi g^{ab} - V[\phi] \bigg]\;. 
\label{actioncol}
\end{eqnarray}
Furthermore, we introduce the potential slow-roll parameters (SRP): 
\beq\label{PSR}
\epsilon_V \equiv \frac{M_P^2}{2} \left( \frac{\partial_\phi V}{V}\right)^2, 
\qquad \delta_V \equiv M_P^2  \left( \frac{\partial_{\phi \phi}^2 V}{V}\right).
\eeq

The slow-roll approximation is valid when $\epsilon_V, \delta_V \ll 1$ and within this approximation, the motion equation for the background field  can be approximated by $3 \mH \phi_0' 
= -a^2 \partial_\phi V$ where $\mH$ is the conformal Hubble factor. Furthermore, during slow-roll inflation the Hubble slow-roll parameter $ \epsilon_H \equiv 1 - 
{\mH'}/{\mH^2}$ is almost equal to the potential slow-roll parameter $\epsilon_V$.

We consider a FRW background space-time with scalar perturbations \footnote{In a previous work \cite{LKL15} we have analyzed the case of tensor perturbations of the metric in the context of the model analyzed in this paper and shown that the corresponding tensor modes are strongly supressed. Therefore, in this paper we only consider scalar pertubations to the metric}. Generically we can write  the line element asssociated to the perturbed metric (in the longitudinal gauge) :
\barr
ds^2 = a^2 (\eta) \big\{  -(1-2\varphi) d\eta^2 + 2 (\partial_i B) dx^i d\eta +
+ [ (1-2\psi) \delta_{ij} + 2 \partial_i \partial_j E] dx^i dx^j \big\}.
\earr

It is convenient to work with the Bardeen potential, defined as $\Phi \equiv \varphi + \frac{1}{a} [a (B-E')]'$ and $\Psi \equiv \psi +\mH (E'-B) $ which are gauge invariant quantities.  Furthermore, the perturbations of the inflaton can be expressed by the gauge-invariant fluctuation of the scalar field $\dphi^{(\textrm{GI})} (\eta,\x) = \dphi + \phi_0' (B-E')$. Thus, within the slow-roll approximation, the first order Einstein-equation can be written as:
\beq\label{25x}
\nabla^2 \Psi +\mu \Psi =  4 \pi G \phi_0' \dphi^{'(\textrm{GI})},
\eeq
where $\mu \equiv \mH^2-\mH'\simeq \epsilon_H \mH^2$. The solution  of Eq. \eqref{25x}  in Fourier space can be expressed as
\beq\label{master0}
\Psi_{\nk} (\eta) \simeq  \sqrt{\frac{\epsilon_H}{2}} \frac{H}{M_P k^2} a 
\dphi'_{\nk} (\eta)^{(\textrm{GI})}\;,
\eeq
where $H$ is the Hubble parameter and $M_P^2\equiv1/8 \pi G$ the reduced Planck mass.
In the semiclassical framework, only the matter fields are quantized, and the self-induced collapse generates the curvature perturbations. Therefore, we consider the quantum theory of  $\dphi (\x,\eta)$ in a curved background described by a quasi-de Sitter space-time. Moreover, it is convenient to  work with the 
rescaled field variable $y=a\dphi$.  
Both the field $y$ and the  canonical conjugated momentum 
$\pi \equiv \partial \delta \mathcal{L}^{(2)}/\partial y' = y'-(a'/a)y=a\dphi'$ are promoted to quantum operators so that they satisfy the following equal time commutator relations :  
$[\hat{y}(\x,\eta), \hat{\pi}(\x',\eta)] = i\delta (\x-\x')$ and 
$[\hat{y}(\x , \eta) , \hat{y}(\x' , \eta)] = 
 [\hat{\pi}(\x , \eta) , \hat{\pi}(\x' ,\eta)] = 0 $ . 
Next, we can expand  the field operator in Fourier modes:
\beq
\hat{y}(\eta,\x) = \frac{1}{L^3} \sum_{\nk} \hat{y}_{\nk} (\eta) e^{i \nk \cdot 
\x},  
\eeq
with an analogous expression for $\hat{\pi}(\eta,\x)$. Note that  the sum is over 
the wave vectors $\vec k$ satisfying $k_i L=2\pi n_i$ 
for $i=1,2,3$ with $n_i$ integer and $\hat y_{\nk} (\eta) \equiv y_k(\eta) 
\ann_{\nk} + y_k^*(\eta) \cre_{-\nk}$ and  $\hat \pi_{\nk} (\eta) \equiv 
g_k(\eta) \ann_{\nk} + g_{k}^*(\eta) \cre_{-\nk}$, with $g_k(\eta) = y_k'(\eta) 
- \mH y_k (\eta)$.  The motion equation of each mode $y_k(\eta)$ reads: 

\beq\label{ykmov2}
y''_k(\eta) + \left(k^2 - \frac{2+3(\epsilon_H - \delta_V)}{\eta^2} \right) 
y_k(\eta)=0.
\eeq

The choice of $y_k(\eta)$ reflects the choice of a vacuum state for the field. In what follows, we proceed as in standard inflationary models and choose the so-called Bunch-Davies vacuum: 

\beq\label{nucolapso}
y_k (\eta) = \left( \frac{-\pi \eta}{4} \right)^{1/2} e^{i[\nu + 1/2] (\pi/2)}
H^{(1)}_\nu (-k\eta), 
\eeq
where $ \nu \equiv 3/2 + \epsilon_H -\delta_V$  and $H^{(1)}_\nu (-k\eta)$ is 
the Hankel function of the first kind of order 
$\nu$. We will not consider the phase $e^{i[\nu + 1/2] (\pi/2)}$ from Eq. \ref{nucolapso} since it has no observational consequence.

Up to  this point the only difference in the treatment of perturbations with standard inflationary models is the semi-classical gravity approach: we only consider at the quantum level the inflaton field perturbations (the metric perturbations remain classic). The collapse hypothesis assumes that  at a certain time $\eta_c^k$ the part of the state characterizing the mode $k$ jumps to a new state, which is no longer homogeneous and isotropic. The collapse is considered to operate similar to a ``measurement", even though there is no external  observer or detector involved. For this, we consider Hermitian operators, which are susceptible of direct measurement in quantum ordinary mechanics.
Therefore, we separate $\hat y_{\nk} (\eta)$ and $\hat \pi_{\nk} (\eta)$ into 
their real and imaginary parts $\hat y_{\nk} (\eta)=\hat y_{\nk}{}^R (\eta) +i 
\hat y_{\nk}{}^I (\eta)$ and $\hat \pi_{\nk} (\eta) =\hat \pi_{\nk}{}^R (\eta) 
+i \hat \pi_{\nk}{}^I (\eta)$, such that the operators $\hat y_{\nk}^{R, I} 
(\eta)$ and $\hat \pi_{\nk}^{R, I} (\eta)$ are  Hermitian operators. Thus, 

\begin{subequations}\label{operadoresRI}
\beq
\hat{y}_{\nk}^{R,I} (\eta) = \sqrt{2} \textrm{Re}[y_k(\eta) 
\hat{a}_{\nk}^{R,I}], 
\eeq
\beq
\hat{\pi}_{\nk}^{R,I} (\eta) = \sqrt{2} 
\textrm{Re}[g_k(\eta) \hat{a}_{\nk}^{R,I}],
\eeq
\end{subequations}
where $\hat{a}_{\nk}^R \equiv (\hat{a}_{\nk} + \hat{a}_{-\nk})/\sqrt{2}$, 
$\hat{a}_{\nk}^I \equiv -i (\hat{a}_{\nk} - \hat{a}_{-\nk})/\sqrt{2}$. 

The commutation relations for the $\hat{a}_{\nk}^{R,I}$ are non-standard:
\beq\label{creanRI}
[\hat{a}_{\nk}^{R,I},\hat{a}_{\nk'}^{R,I \dag}] = L^3 (\delta_{\nk,\nk'} \pm 
\delta_{\nk,-\nk'}), 
\eeq
where the $+$ and the $-$ sign corresponds to the commutator with the $R$ and 
$I$ labels respectively; all other commutators vanish.
It is also important to emphasize that the  vacuum state   defined  by $ \ann_{\nk}{}^{R,I} |0\ket =0$  is  fully translational  and rotationally invariant (see the formal proof in 
Appendix A of Ref. \cite{LLS13}).

To connect the quantum theory of the inflaton perturbations with the primordial curvature perturbation, we choose  to 
work in the longitudinal gauge. We write Eq. \eqref{master0}  in terms of  the expectation value of the conjugated momentum
\beq\label{masterpi}
\Psi_{\nk} (\eta) \simeq  \sqrt{\frac{\epsilon_H}{2}} \frac{H}{M_P k^2}  \bra 
\hat{\pi}_{\nk} (\eta) \ket \;.
\eeq

It follows from the  above equation that in the vacuum state $ \bra \hat{\pi}_{\nk} (\eta) \ket =0$, 
which implies $\Psi_{\nk} =0$, i.e., there are no perturbations of the symmetric 
background space-time. It is only after the collapse has taken place 
($|\Theta\ket \neq |0\ket$)  that $\bra \hat{\pi}_{\nk} (\eta) \ket_\Theta \neq 
0$ generically and $\Psi_{\nk} \neq 0$; thus, the primordial inhomogeneities 
and anisotropies arise from the quantum collapse. Next, we need to  specify the dynamics of the expectation values $\bra  \hat{y}^{R, I}_{\nk} 
(\eta)   \ket$ and $\bra  \hat{\pi}^{R, I}_{\nk} (\eta)   \ket$, evaluated in  the post-collapse state, which will depend on the expectation values evaluated at the time of collapse of each mode of the field $\tc$.

\subsection{Collapse schemes}\label{esquemas-colapso}

Even though a full workable relativistic collapse mechanism is still unknown, some relativistic collapse mechanism have been recently  proposed \cite{bedingham2010,pearle2014}. 
On the other hand, some non-relativistic objective collapse models have been  studied previously in the  literature \cite{ghirardi1985,pearle1989,bassi2003,bassi2013}.
In this paper, we will not consider a specific collapse mechanism. Instead, we will follow the approach of Ref.\cite{LLP15} and assume that whatever the collapse mechanism is behind, after the  collapse, the expectation values of the field and momentum operators  in each mode  will  be related to the uncertainties  of the  initial state. 
We  could   consider  various  possibilities  for such relations, e.g., different {\it collapse schemes}. 

In this work, we focus on the \emph{Newtonian} and \emph{Wigner} schemes studied in Ref.\cite{LLP15}. 
We do not consider the {\it independent } scheme studied in the same work since it has been shown that its CMB angular spectrum is indistinguishable from the prediction of the standard inflationary model.

\subsubsection{Newtonian collapse scheme}

In this scheme the collapse affects only the conjugated momentum variable, i.e.,  

\beq\label{esquemanewt}
 \bra \hat{y}^{R,I}_{\nk}(\eta^c_{\nk})\ket  = 0, \qquad
  \bra \hat{\pi}^{R,I}_{\nk}(\tc) \ket = x_{\nk,2}^{R,I}
  \sqrt{\left(\Delta \hat{\pi}^{R,I}_{\nk} (\tc) \right)^2_0}.
\end{equation}
where, $x_{\nk,2}^{(R,I)}$ represents a random Gaussian 
variable normalized and centered at zero and  the quantum uncertainties 
can be expressed as

\barr
\left(\Delta \hat{y}^{R,I}_{\nk} (\tc) \right)^2_0 &=& \frac{L^3 \pi |z_k|}{16 k} \left[ J_\nu^2 (|z_k|) + Y_\nu^2 (|z_k|) 
\right], 
\earr

\barr
\left(\Delta \hat{\pi}^{R,I}_{\nk} (\tc) \right)^2_0 &=&\frac{L^3 \pi k }{16} \nonumber 
\times \bigg[ \left( \frac{-\alpha 
J_\nu 
(|z_k|)}{\sqrt{|z_k|}} + \sqrt{z_k|} J_{\nu+1} (|z_k|) \right)^2 \nonumber \\
&+& \left( \frac{-\alpha Y_\nu (|z_k|)}{\sqrt{|z_k|}} + \sqrt{|z_k|} Y_{\nu+1} 
(|z_k|) \right)^2 \bigg], \nonumber \\
\earr
where $J_\nu $  and $Y_\nu$ are the Bessel functions of the first and second kind respectively; $z_k \equiv k \tc$ and $\tc$ is the time of collapse for each mode.

\subsubsection{Wigner collapse scheme}

This scheme is 
motivated by considering the correlation between $\hat{y}^{R,I}$ and
$\hat{\pi}^{R,I}$ existing in the pre-collapse state and characterize it in terms of the Wigner function. The Wigner function of the vacuum state is a bi-dimensional Gaussian function. Thus, in this scheme in the post-collapse state  the expectation value of the fields will be characterized by
\begin{subequations}\label{esquemawig}
 \beq
\bra \hat{y}^{R,I}_{\nk}(\eta^c_{\nk})\ket  = x_{\nk}^{R,I} \Lambda_k (\tc) 
\cos \Theta_k (\tc), 
\eeq
\beq
  \bra \hat{\pi}^{R,I}_{\nk}(\tc) \ket = x_{\nk}^{R,I} k \Lambda_k (\tc) 
\sin \Theta_k (\tc),
\eeq
\end{subequations}
where $x_{\nk}^{R,I}$ is a random variable, characterized by a Gaussian 
probability distribution function  centered at zero with  spread one. The 
parameter $\Lambda_k (\tc)$ represents the major semi-axis of the ellipse 
characterizing the bi-dimensional Wigner function that can be considered a 
Gaussian in two dimensions.  $\Theta_k 
(\tc)$ is the angle between that axis and the $\hat{y}_{\nk}^{R,I}$ axis. For  
details involving the Wigner function and the collapse scheme we refer the reader to  
 Ref. \cite{US08}.From Ref.\cite{LLP15}, we can also write the expression for $\Lambda_k$
and $\Theta_k $:
\barr\label{lambdak}
&\Lambda_k& = (2L)^{3/2} \sqrt{\frac{\pi |z_k|}{4k}} \left[ J_\nu^2 (|z_k|) + 
Y_\nu^2 (|z_k|) \right]^{1/2} \bigg[ S(|z_k|)  \nonumber \\
&-& \sqrt{S^2 (|z_k|) -  \left(\frac{\pi 
|z_k|}{2}\right)^2    (J_\nu^2 (|z_k|) + Y_\nu^2 (|z_k|) )^2} \bigg]^{-1/2}, 
\nonumber \\
\earr

\barr\label{2thetak}
&\tan 2\Theta_k& =  -\frac{\pi^2 |z_k|}{4}  \left[ J_\nu^2 (|z_k|) + Y_\nu^2 
(|z_k|) \right] \nonumber \\
&\times& \left[  S(|z_k|) -  \frac{\pi |z_k|}{8} \left( 
J_\nu^2 (|z_k|) 
+ 
Y_\nu^2 (|z_k|) \right)^2 \right]^{-1} \nonumber \\
&\times& [ -2\nu \left( J_\nu^2 (|z_k|) + Y_\nu^2 (|z_k)  \right) + |z_k| 
\nonumber \\
&\times& \left( J_\nu (|z_k|) J_{\nu+1} (|z_k|) +  Y_\nu (|z_k|) Y_{\nu+1} 
(|z_k|)       
\right)  ], \nonumber \\  
\earr
where

\barr\label{defS}
&S(|z_k|)& \equiv 1 + \frac{\pi^2}{16} \bigg\{  |z_k|^2 (J_\nu^2 (|z_k|) + 
Y_\nu^2 (|z_k|))^2 
\nonumber \\ 
&+& 4 \bigg[ J_\nu^2 (|z_k|) + Y_\nu^2 (|z_k|) - |z_k| ( J_\nu (|z_k|) 
J_{\nu+1} 
(|z_k|) \nonumber \\ 
&+& Y_\nu (|z_k|) Y_{\nu+1} (|z_k|) )   \bigg]^2  \bigg\}. 
\nonumber \\
\earr

\section{Primordial Power Spectrum for the collapse models}
\label{primordialpower}

In this section we briefly review the procedure to obtain the primordial power spectra for the collapse models and show examples for some specific values of the collapse parameters. Furthermore, we show the predictions for the proposed $\eta_c^k$ parametrisation on the current observables.

Let us introduce how the temperature anisotropies $\Theta (\hat {n}) \equiv \delta 
T/T_0$ of the CMB can be connected with the parameters characterizing the collapse. The coefficients $a_{lm}$ of the spherical harmonic expansion of  $\delta T/T_0$ are 

\beq\label{alm0}
a_{lm} = \int \Theta (\hat n) Y_{lm}^\star (\theta,\varphi) d\Omega,
 \eeq
with $\hat n = (\sin \theta \sin \varphi, \sin \theta \cos \varphi, \cos 
\theta)$ and $\theta,\varphi$ the coordinates on the celestial two-sphere.  We use a Fourier decomposition for the temperature anisotropies $\Theta (\hat n) = \sum_{\nk} \frac{\Theta (\nk)}{L^3} e^{i \nk \cdot R_D \hat n}$ with $R_D$ being the radius of the last scattering surface. Furthermore $\Theta (\nk) = T(k) \mR_{\nk}$, where the  initial curvature perturbation $\mR_{\nk}$ is connected to the temperature anisotropies $\Theta (\nk)$ by the transfer function $T(k)$ which contains the physics between the radiation era and the present. Consequently, the coefficients $a_{lm}$, in terms of the modes $\mR_{\nk}$, are  given by 
\beq\label{alm2}
a_{lm} = \frac{4 \pi i^l}{ L^3} \sum_{\nk} j_l (kR_D) Y_{lm}^\star(\hat k) T 
(k) 
\mR_{\nk},
\eeq
with $j_l (kR_D)$ being the spherical Bessel function of order $l$.

It has been shown \cite{US08,LLP15} that the the coefficients $a_{lm}$ are directly related 
to the random variables $x_{\nk}$. Therefore, the coefficients $a_{lm}$ are 
in effect a sum  of  random complex  numbers like an effective  two-dimensional  random walk. On the other hand, one 
cannot give a perfect estimate for the direction of the final displacement  
resulting from the random walk, but  one might give an estimate for 
the length of the displacement. Thus, we can make an estimate for the most likely value of  $|a_{lm}|^2$ and interpret it as the theoretical prediction for the observed value. Furthermore,  since the collapse is being 
modeled by a random process, we can consider a set of possible realizations of 
such process characterizing the universe in an unique manner, i.e.,  
characterized by the random variables $x_{\nk}$. If the probability 
distribution function of $x_{\nk}$ is Gaussian, then we can identify the most likely value 
$|a_{lm}|^2_{\text{ML}}$ with the mean value $\overline{|a_{lm}|^2}$ of all 
possible realizations, i.e., $|a_{lm}|^2_{\text{ML}}= 
\overline{|a_{lm}|^2}$. 
Furthermore, the quantity that is used in the statistical analysis to compare with observational data is the angular power spectrum: $C_l = (2l+1)^{-1} \sum_m |a_{lm}|^2$. Therefore, we can use the prediction for $|a_{lm}|^2_{\text{ML}}$ for each collapse scheme 
to give a theoretical prediction for the $C_l's$:

\beq\label{clcolapso}
C_l = { 4 \pi} \int_0^\infty \frac{dk}{k} j_l^2 (kR_D) T(k)^2 
\frac{\mathcal{C}}{\pi^2} Q(|z_k|) k^{3-2\nu},
\eeq
where
\beq\label{C}
\mathcal{C}\equiv \frac{\pi}{ M_P^2 \epsilon_H} \left( 2^{\nu-11/2} 
\Gamma(\nu-1) H |\eta|^{3/2-\nu}    \right)^2,
\eeq
and we have taken the limit $L\to \infty$ and $\nk \to$ continuum in order to 
go from sums over discrete $\nk$ to integrals over $\nk$. The function 
$Q(|z_k|)$ varies for each collapse scheme (see Ref.\cite{LLP15}) and depends on the collapse time of each mode through $z_k$.
On the other hand, the time of collapse can happen at any time during the 
inflationary regime. In particular, it can occur when the proper wavelength of 
the mode is bigger or smaller than the Hubble radius.  In this paper, we focus on the case where the proper wavelength associated to the mode is smaller 
than the Hubble radius, at the time of collapse, then $k \gg a(\tc) H$, which 
is equivalent to $-k \tc \gg 1$. The approximated collapse power spectrum, 
when $-k \tc = |z_k| \to \infty$, is given by \cite{LLP15}

 \beq\label{pscolapsozinfty}
P(k) \simeq \frac{\mathcal{C}}{\pi^2} \Upsilon (|z_k|) k^{n_s-1},
\eeq
Taking $\nu=2 - \frac{n_s}{2}$, for each scheme the function $\Upsilon(|z_k|)$ is

\begin{subequations}\label{pdentro3esquemas}

\barr
& & \Upsilon(|z_k|)^{\textrm{newt}} \equiv \frac{4}{\pi^2} \nn
&\times& \left[  1 + 
\frac{1}{|z_k|^2} \left( -2\nu + \frac{\Gamma(\nu + 5/2)}{2 \Gamma(\nu + 1/2)}  
\right)^2       \right] \nonumber \\
&\times& \left[ \cos \beta(\nu,|z_k|) - \frac{\sin \beta(\nu, |z_k|) }{2|z_k|} 
\frac{\Gamma(\nu+3/2)}{\Gamma(\nu - 1/2)}     \right]^2,
\label{pnewtdentro}
\earr
\barr
& & \Upsilon(|z_k|)^{\textrm{wig}} \equiv \frac{16}{\pi^2} \bigg\{  \bigg[ 
\frac{2\nu}{\zk^{3/2}} \nn
&\times& \left( \cos \beta(\nu,|z_k|) - \frac{\sin \beta(\nu, 
|z_k|) }{2|z_k|} \frac{\Gamma(\nu+3/2)}{\Gamma(\nu - 1/2)}     \right)   
\nonumber \\
&-& \left( \sin \beta(\nu,|z_k|) + \frac{\cos \beta(\nu, |z_k|) }{2|z_k|} 
\frac{\Gamma(\nu+5/2)}{\Gamma(\nu + 1/2)}     \right)            \bigg] \cos 
\Theta_k   \nonumber \\
&+&  \left[ \cos \beta(\nu,|z_k|) - \frac{\sin \beta(\nu, |z_k|) }{2|z_k|} 
\frac{\Gamma(\nu+3/2)}{\Gamma(\nu - 1/2)}     \right] \sin \Theta_k  \bigg\}^2, 
\nn
\label{pwignerdentro}
\earr
\end{subequations}
where $\beta(\nu,|z_k|) \equiv |z_k| - (\pi/2) (\nu+1/2)$ and $\tan 2\Theta_k 
\simeq -4/3\zk$.

It follows from Eq. \ref{pscolapsozinfty}  that if we  consider $z_k$ equal to a constant
we recover the dependence in $k$ of the standard model. Furthermore, in previous works  \cite{LSS12,LLP15}, small departures from this expression 
were considered. In this work, we go one step further and consider a different 
$z_k = \mathcal{A} + \mathcal{B}/k$, which implies the following expression for the collapse time of each mode
\beq
\eta_c^k = \frac{\mathcal{A}}{k}+\frac{\mathcal{B}}{k^2}, 
\label{ctime}
\eeq
where \A is adimensional and \B has units of {\rm Mpc}$^{-1}$. 
Here we mention that the inflationary expansion period 
corresponds to negative conformal time, so we choose to work with negative values for \A and \B.
Note that, differently from the previous expressions studied in Refs.~\cite{LSS12,LLP15}, the above parameterisation predicts a primordial power spectrum which is significantly different from the standard prediction over the entire multipole interval and, in particular, at low-$\ell$, providing also a possible explanation for  the observed lack of power in the CMB temperature anisotropies at large angular scales~\cite{Ade:2015xua, Copi:2013cya}.
\begin{figure*}[t]
	\centering
	\includegraphics[width=0.3\hsize]{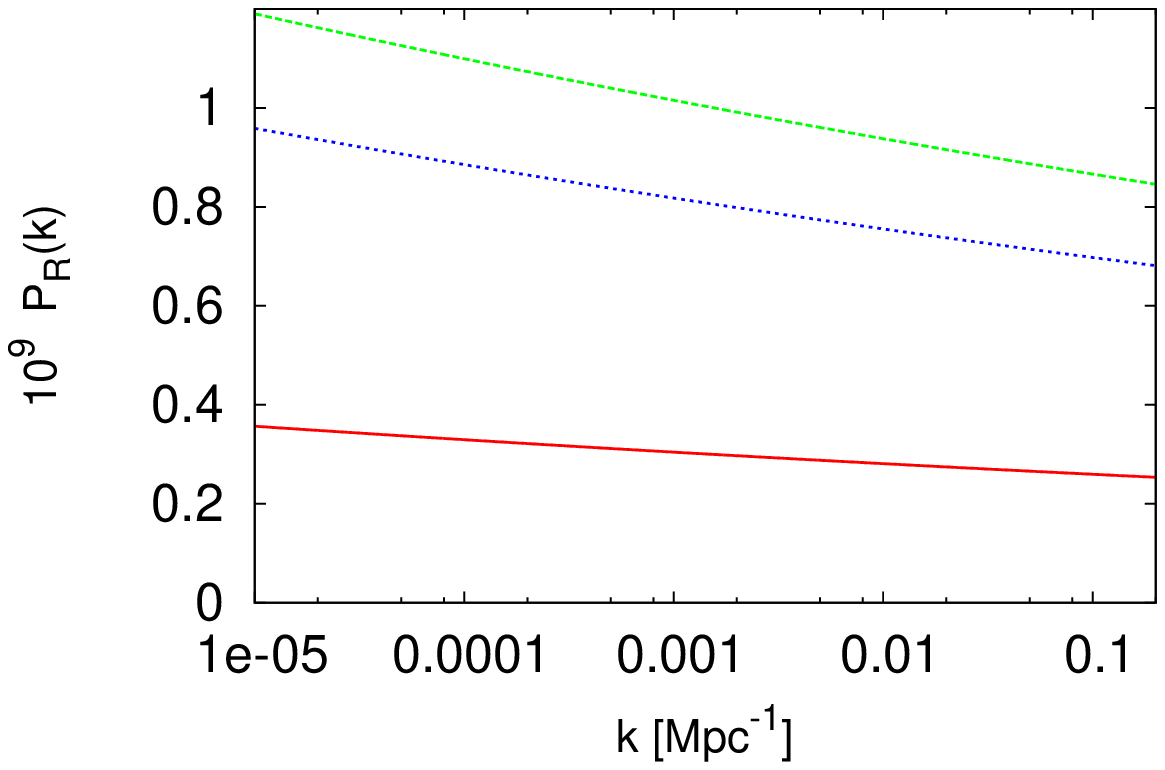}
    \includegraphics[width=0.3\hsize]{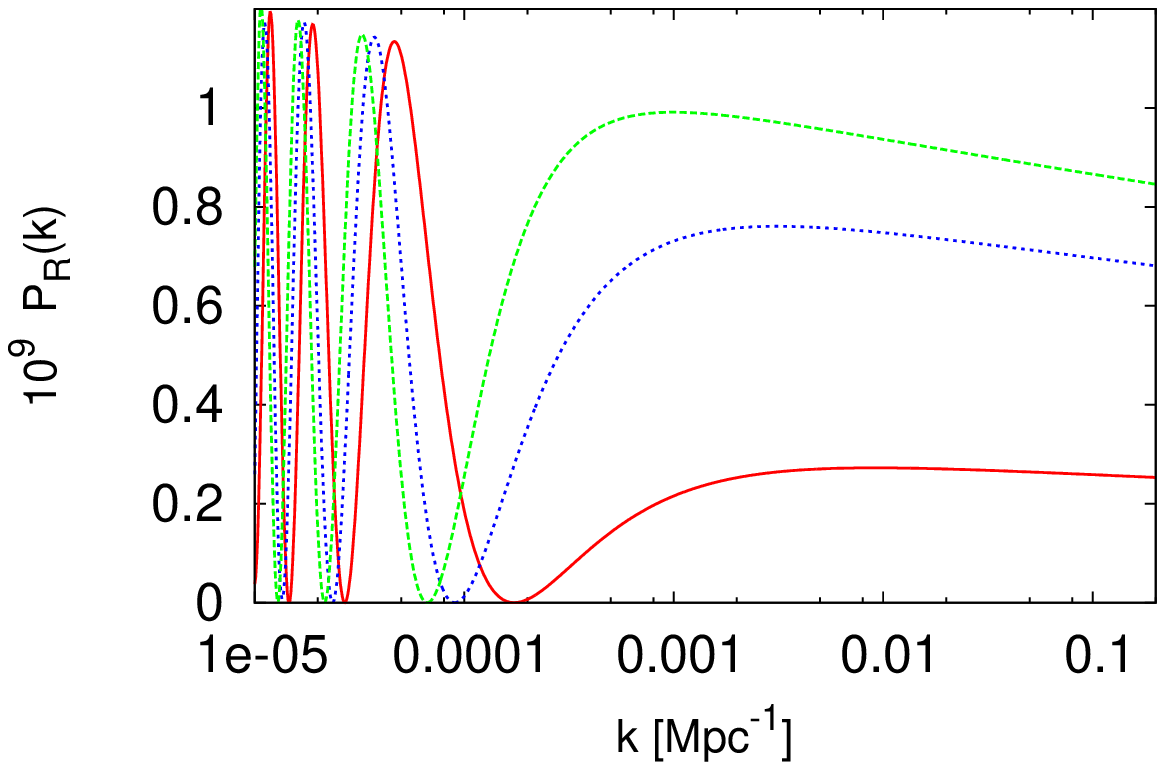}
	\includegraphics[width=0.3\hsize]{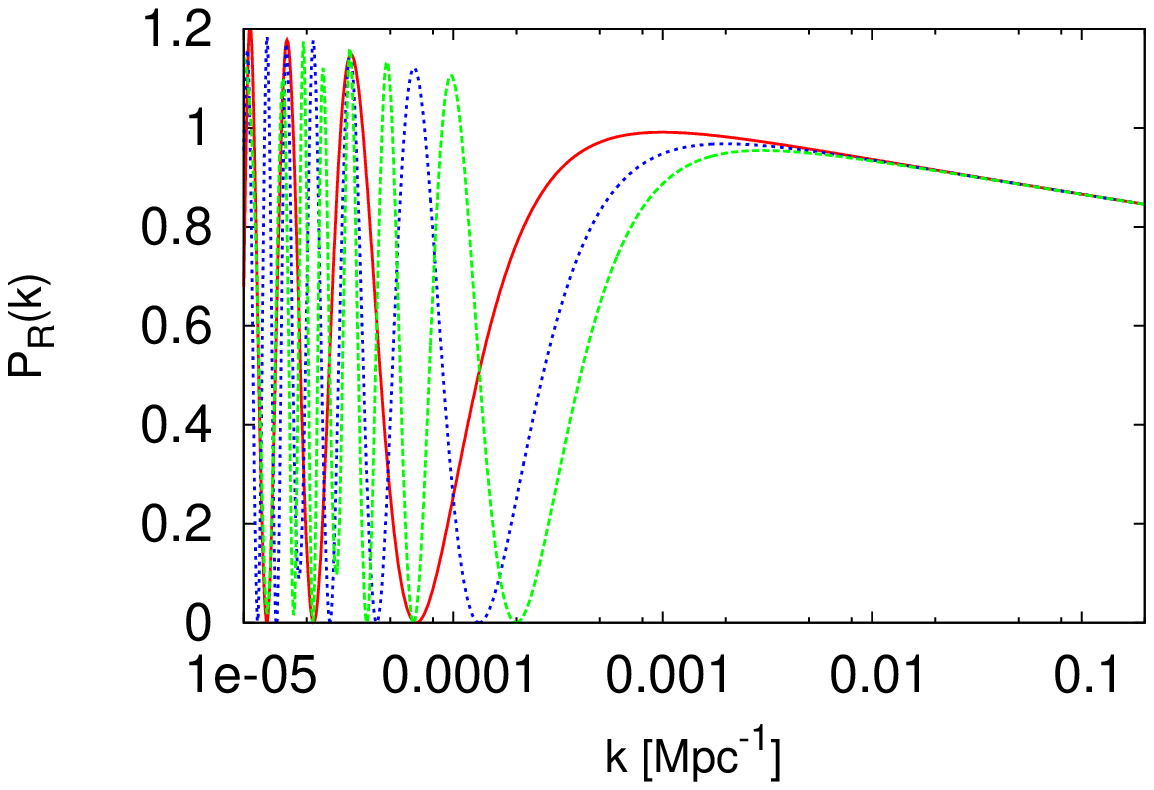}
	\includegraphics[width=0.3\hsize]{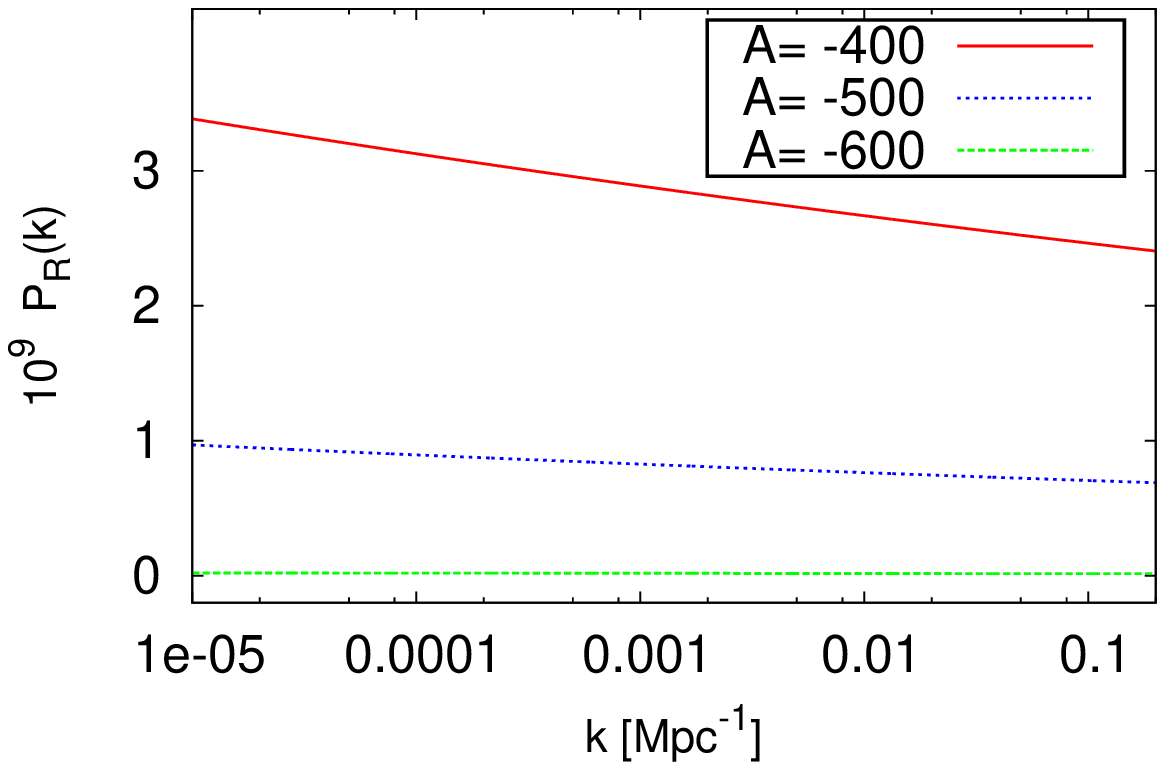}
	\includegraphics[width=0.3\hsize]{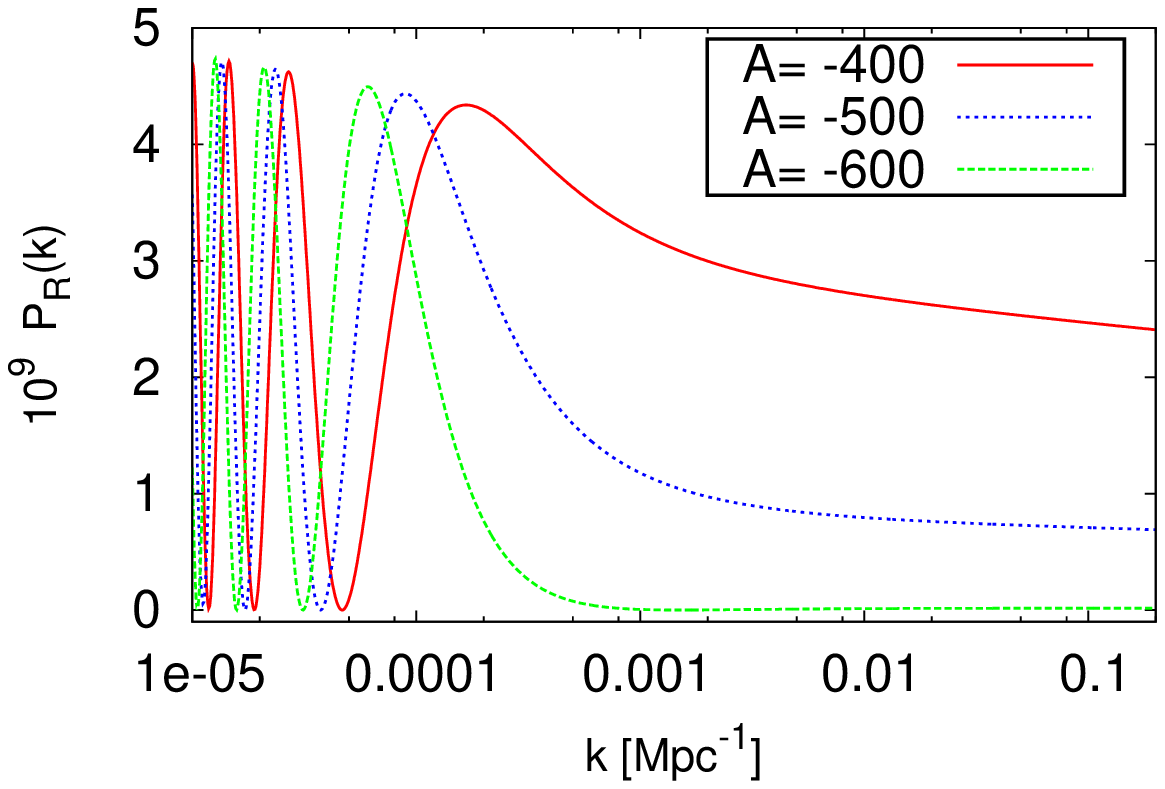}
	\includegraphics[width=0.3\hsize]{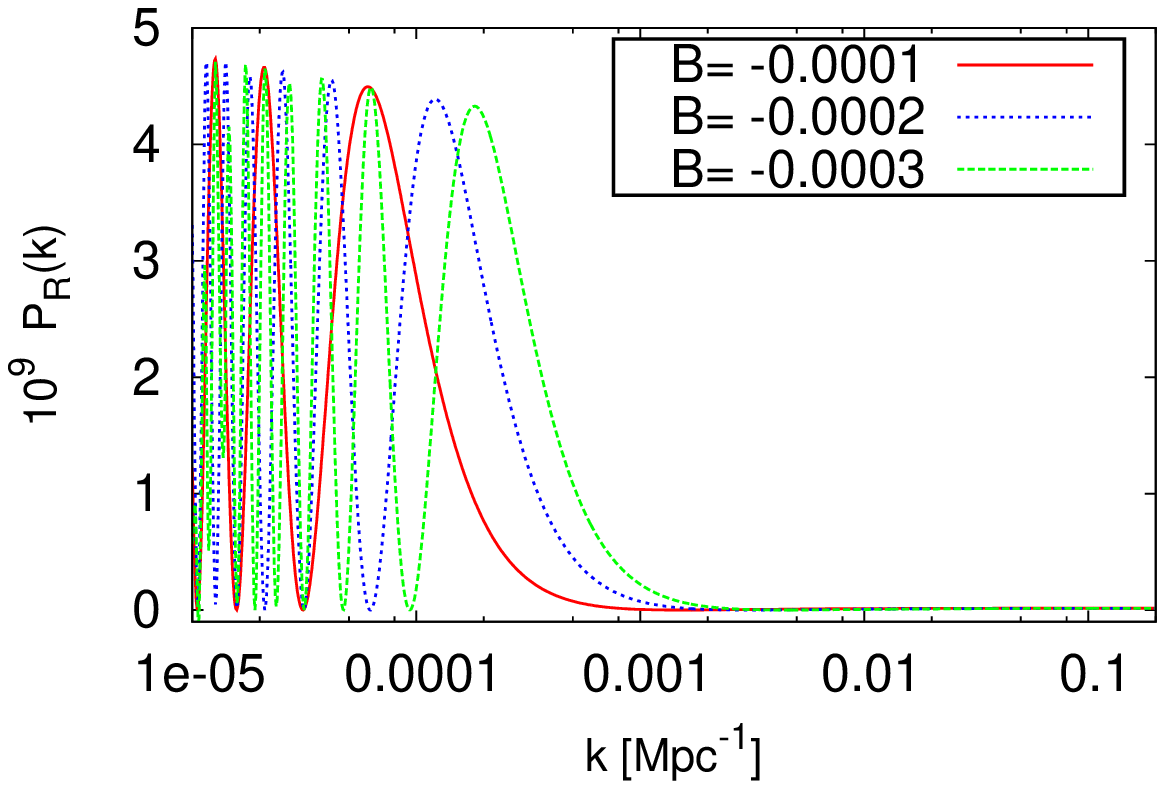}
	\caption{Primordial power spectra for the {\it Newtonian} collapse (top) and {\it Wigner} collapse (bottom) scheme parameterisations. 
For both models the label is reported in the bottom line	.
Left: the collapse  parameter \B is fixed to zero, while \A assumes different values. 
Middle: the same as in the previous panel for $\mathcal{B}= - 0.0001$. Right: the collapse parameter \A is fixed at $\mathcal{A}= - 600$, while \B assumes different values.}
	\label{fig:PR}
\end{figure*}
\begin{figure}
	\centering
	\includegraphics[width=0.6\hsize]{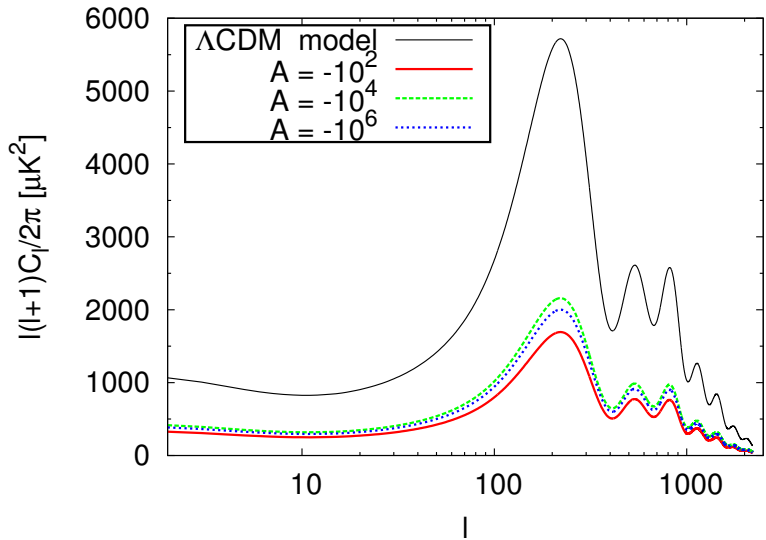}
	\caption{Anisotropy power spectrum for the {\it Newtonian} scheme model using different values of $\mathcal{A}$, while $\mathcal{B}$ is fixed to zero. The black line stands for the $\Lambda$CDM model.}
	\label{fig:ClsB0}
\end{figure}
\begin{figure}
	\centering
	\includegraphics[width=0.4\hsize]{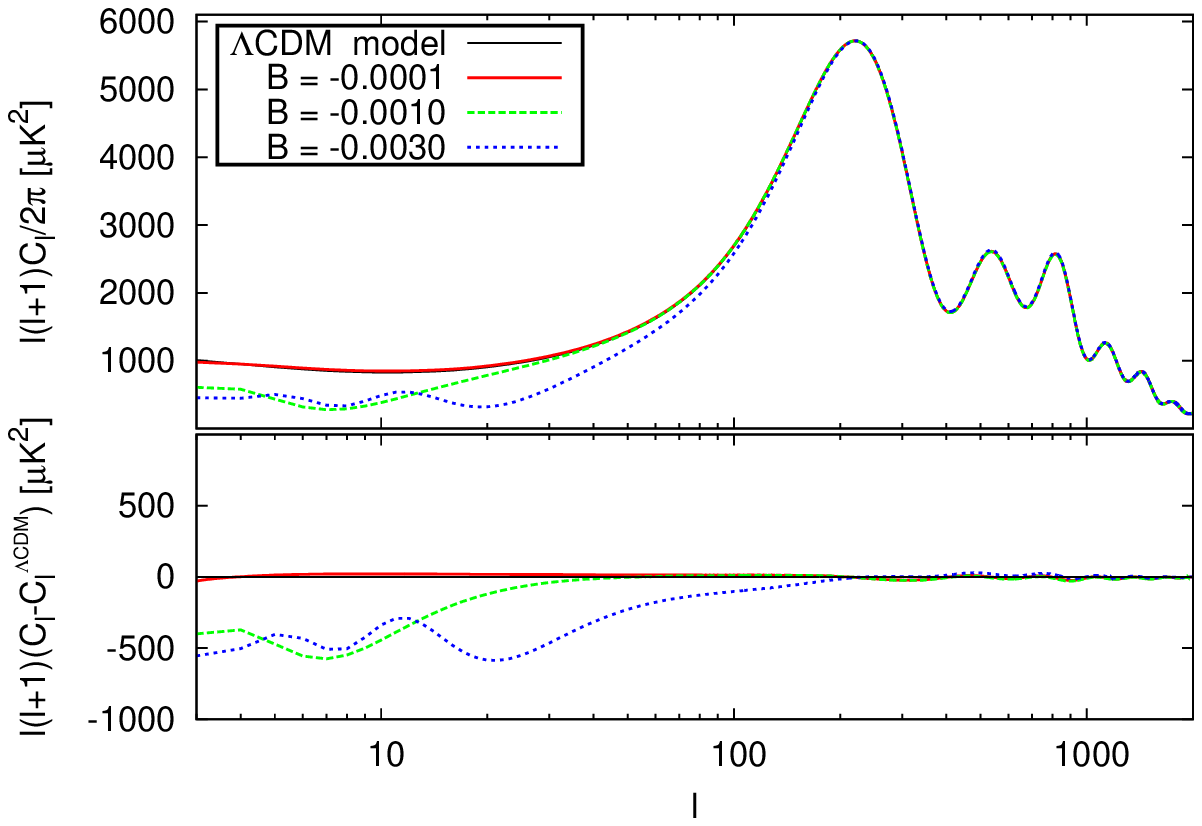}
	\includegraphics[width=0.4\hsize]{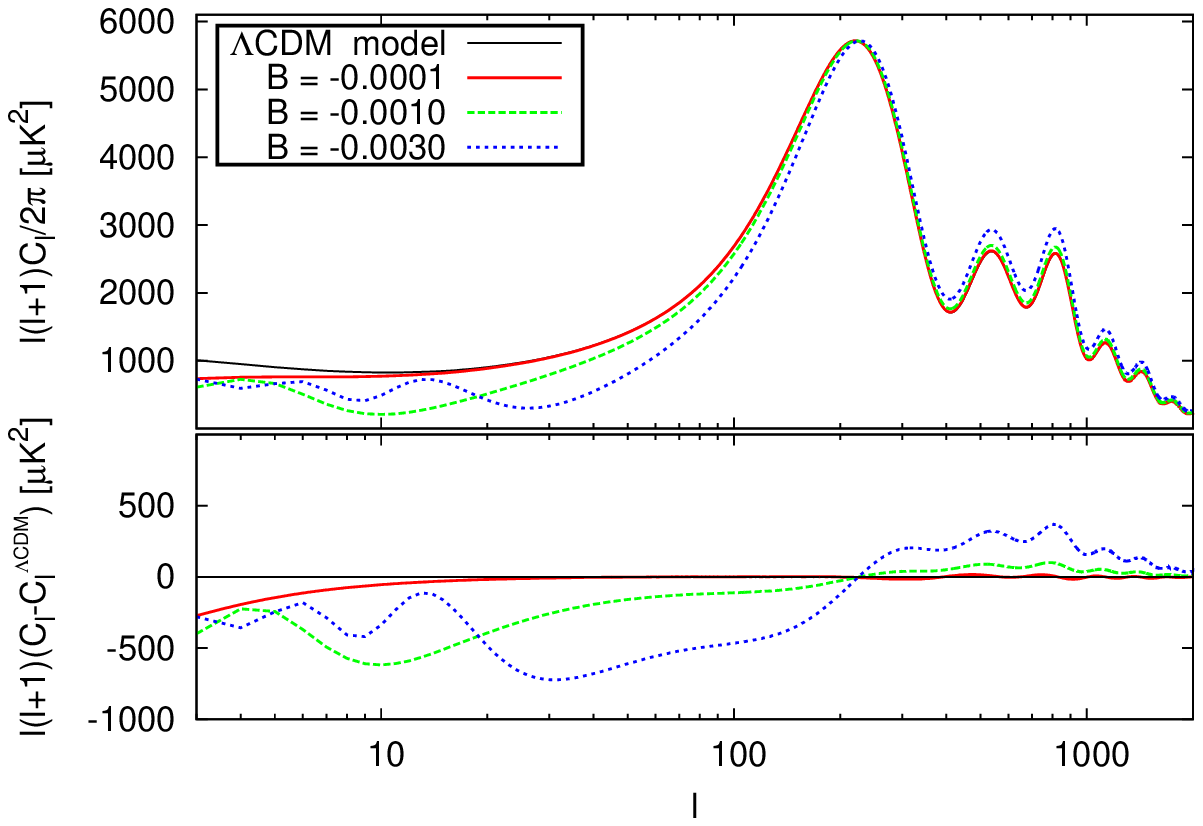}
	\caption{Anisotropy power spectrum and differential plot with respect to the $\Lambda$CDM model for the {\it Newtonian} scheme model (left) using $\mathcal{A}=-600$ and different values of $\mathcal{B}$; for the {\it Wigner} scheme model (right) using $\mathcal{A}=-800$ and the same values of $\mathcal{B}$ values of the previous panel. The black line stands for the $\Lambda$CDM model.}
	\label{fig:ClsNewtWigner}
\end{figure}

The effect on the predicted primordial power spectrum 
using both the {\it Newtonian} collapse and {\it Wigner} collapse schemes is showed in  Fig. (\ref{fig:PR}). It is found to be essentially a power-law with superimposed oscillations due to the \B parameter. Indeed, the parameter \A determines the amplitude of the spectrum (see the left and middle panels), while the parameter \B influences the oscillations frequency (see the right panel) and determines the scale where the spectrum recovers the familiar $k^{n_s-1}$ form of the standard scenario. We note the difference in amplitude oscillation between the two schemes: using the same values of \A and \B the {\it Newtonian} collapse curves oscillates between [$0:1.2$] while the {\it Wigner} collapse curve covers the amplitude range [$0:5$]. Finally, we note the different behavior between the two parameterisations in the middle panel: while the {\it Newtonian} case approaches the power-law behavior from low values of the primordial spectrum, the {\it Wigner} collapse scheme behaves the other way around. This is due to the particular choice of the \A value, i.e. the{\it  Wigner} scheme recovers the same {\it Newtonian} shape for $A=-800$.

Next, we show the effects of assuming the collapse hypothesis  for both the schemes considered in this paper  on the CMB temperature auto-correlation angular power spectra. 
When the collapse parameter $\mathcal{B}=0$, as discussed before, we recover the standard $\Lambda$CDM model prediction unless a normalization factor. 
As a consequence the collapse parameter $\mathcal{A}$ will be highly degenerate with the scalar amplitude $A_s$. Furthermore, Figure (\ref{fig:ClsB0}) shows that the amplitude factor varies with  different values of the collapse parameter $\mathcal{A}$ and we verified  that the curves overlap within a normalization.  It is worth mentioning that such a  variation is highly non-linear.  

In Figure (\ref{fig:ClsNewtWigner}) we show the temperature auto-correlation (TT) power spectrum for a fixed value of $\mathcal{A}$ and different values of the collapse parameter $\mathcal{B}$. The value of $A_s$ is settled in each case in order to match the maximum of the first Doppler peak with the reference model one, namely the best fit $\Lambda$CDM model obtained by the Planck collaboration  \cite{Ade:2015xua}.  We can see that the value of the collapse parameter $\mathcal{B}$ affects the low multipole region. Besides,  as  the absolute value of $\mathcal{B}$ increases,  the change takes the form of oscillations while as the value of $\mathcal{B}$ approaches to $0$ the shape of the spectrum approaches to the reference model one. 
Furthermore, we also analyzed the $EE$ and $TE$ angular power spectra and found only a tiny variation at low multipoles with respect to the reference model. 
On the other hand, for the {\it Wigner} scheme, we note a  change in the height of the secondary peaks for increasing values of $\mathcal{B}$.  This is due to a more sensitivity with the variations in the collapse parameter $\mathcal{B}$ of the latter with respect to the {\it Newtonian} scheme. In other words, the {\it Newtonian} scheme shows the same behaviour when bigger values of \B are assumed. In agreement with the growth in high multipoles for increasing values of $\mathcal{B}$, we find small increases in the  $EE$ and $TE$ power spectra.

\section{Analysis Method}
\label{Method}

In order to compute the CMB anisotropies spectrum for given values of the collapse schemes parameters, we use a modified version of the {\sc CAMB}~(\cite{camb}) code to include the primordial power spectrum of the collapse models. We perform a Monte Carlo Markov chain analysis using the available package {\sc CosmoMC}~\cite{Lewis:2002ah} and implement the nested sampling algorithm of {\sc Multinest} code~\cite{Feroz:2008xx,Feroz:2007kg,Feroz:2013hea} to obtain our results and calculate the Bayesian evidence factor.  
In our Bayesian analysis we use the most accurate Importance Nested Sampling (INS)~\cite{Cameron:2013sm, Feroz:2013hea} instead of the vanilla Nested Sampling (NS), requiring a INS Global Log-Evidence error of $\leq 0.02$ .

\begin{figure}
	\centering
		\includegraphics[width=0.9\hsize]{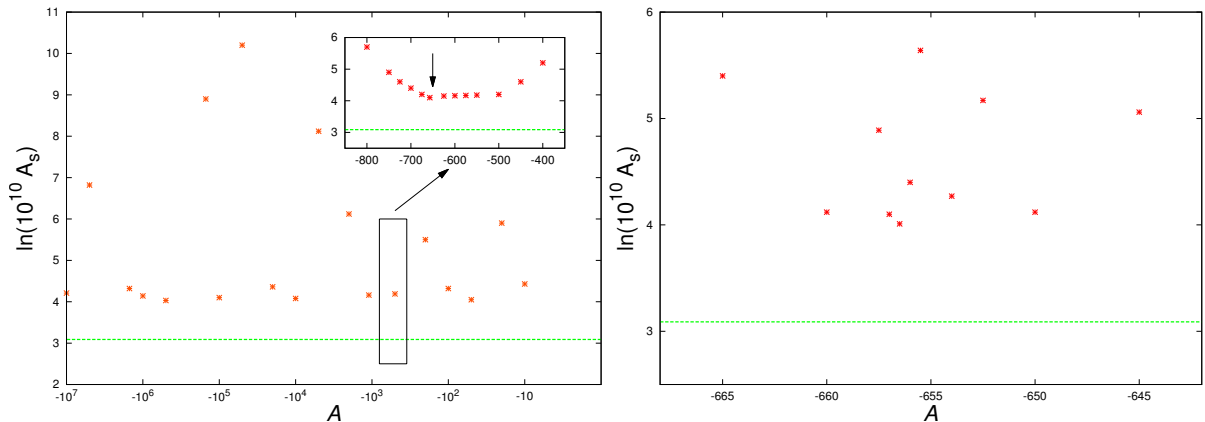}	
	\caption{Behaviour of the collapse parameter \A with the primordial scalar amplitude $A_s$ for the {\it Newtonian} scheme model. In the left we show the relation in the \A range [$-10^{-7}$ : $-10$] and the zoom in the \A range [$-800$ : $-400$]. In the right we plot the progressive zoom in the \A range [$-665$ : $-645$]. The reference green line draws the $\Lambda$CDM best fit value. }
	\label{fig:A_As_newtonian}
\end{figure}
We consider extensions of the minimal $\Lambda$CDM model, adding the collapse schemes parameters \A and \B to the usual set of cosmological parameters: the baryon density, $\Omega_bh^2$, the cold dark matter density, $\Omega_ch^2$, the ratio between the sound horizon and the angular diameter distance at decoupling, $\theta$, the optical depth, $\tau$, the primordial scalar amplitude, $A_s$, and the primordial spectral index $n_s$. We consider purely adiabatic initial conditions, fix the sum of neutrino masses to $0.06~eV$ , and limit the analysis to scalar perturbations with $k_0=0.05$ $\rm{Mpc}^{-1}$. We also vary the nuisance foregrounds parameters~\cite{Aghanim:2015xee}.

In our analysis we use the CMB data set from the latest Planck Collaboration release~\cite{Ade:2015xua},  considering the high-$\ell$ Planck temperature data (in the range of $30< \ell <2508$) from the 100-,143-, and 217-GHz half-mission T maps, and  the low-P data by the joint TT,EE,BB and TE likelihood (in the range of $2< \ell <29$).
High-$\ell$ polarization data are not used since, as shown in the previous section, the analyzed collapse time expression affects only low multipoles, in both temperature and polarization spectra, recovering the $\Lambda$CDM behavior at small scales.

We work with flat priors for the cosmological parameters, and assume sharp prior intervals for the  collapse parameter \A. 
As seen in Figs. (\ref{fig:PR}) and (\ref{fig:ClsB0}) , the \A parameter just sets the primordial spectrum amplitude, which means that it is highly degenerate with the $A_s$ parameter. Furthermore, the spectrum amplitude value is significantly sensitive to variations of \A. In Fig. (\ref {fig:A_As_newtonian}) we show a progressive zoom in the range values of \A, with respect to the $A_s$ parameter for the {\it Newtonian} scheme. Each point in the plot represents the combination of these parameters that assures the best fit to the current data. We can see that the $A_s$ value oscillates even with a smaller variation of \A, generating significant computational problems.  Testing several values for the \A parameter, we select an interval of values which satisfies the condition for the conformal collapse time $\eta_c^k < 0$ and minimizes the variation of the $A_s$ from the $\Lambda$CDM model value. In the same fashion, we perform the selection of the \A value for the {\it Wigner} scheme (not shown in the figure). We also limit \B into small values, since we also have a particular interest in studying features at low multipoles. We consider the interval of values shown  in Tab. \ref{tab:priors}, however, worth mentioning that we have also explored different ranges of  \B values, e.g.,  $\lesssim \mathcal{B} = -3 \times10^{-3}$, which are strongly ruled out by the current data. Indeed, the increased sensitivity of the {\it Wigner} scheme discussed in the previous section is the reason why we considered a more stringent  prior on $\mathcal{B}$ for this model.

In order to make an appropriate comparison between the collapse model and the standard $\Lambda{CDM}$ model predictions for the CMB angular power spectrum, we use the Bayesian model comparison, that is a very powerful tool to reward the models that fit well the data exhibiting strong predictivity, while models with a large number of free parameters, not required by the data, are penalised for the wasted parameter space. The Bayesian \textit{evidence} $\mathcal{E}$ is defined as the marginal likelihood for the model $M_i$:
%
%
%
\begin{table}[]
    \centering
    \begin{tabular}{|c|ll|}
        \hline
        & & \\
        Parameter   & Newtonian scheme    &  Wigner scheme\\
        \hline  \hline
        & & \\
        \A       & [$-656.99$ : $-656.93$]  &   [$-800.05$ : $-800.00$] \\
        & &\\
        \B  &  [$-5 \times10^{-4}$ : $0$]    & [$-2.5 \times10^{-4}$ : $0$]\\
        & & \\
        \hline  
    \end{tabular}
    \caption{\label{tab:priors} Priors on the {\it Newtonian} and {\it Wigner} collapse scheme parameters considered in the analysis.}
\end{table}
%
%
\begin{equation}
\mathcal{E}_{M_i} =\int d\theta p(x|\theta, M_i) \pi(\theta | M_i)\;.
\end{equation}
where $x$ stands for the data, $\theta$ is the parameters vector and $\pi(\theta|M_i)$ the prior probability distribution function. The ratio of the Bayesian evidence of the two models (the so-called \textit{Bayes Factor}) can be defined as:
\begin{equation}
\mathit{B}_{ij}= \frac{\mathcal{E}_{M_i}}{\mathcal{E}_{M_j}}\;,
\end{equation}
where $M_j$ is the reference model.
The more complicate model $M_i$ (e.g., the one with more parameters with respect to the reference model) inevitably leads to a higher likelihood, but the evidence will favor the simpler model if the fit is as nearly as good, through the smaller prior volume. We assume uniform (and hence separable) priors in each parameter, such that we can write $\pi(\theta|M) = (\Delta \theta_1 ~. . .~\Delta\theta_n)^{-1}$ and
\begin{equation}
\mathit{B}_{ij}= \frac{ \int d\theta p(x|\theta,M_i)}{ \int d\theta' p(x|\theta',M_j)}
\frac{(\Delta \theta_1 ~. . .~\Delta\theta_{n_i})}{(\Delta \theta'_1 ~. . .~\Delta\theta'_{n_j})}\;.
\label{eq:bayes}
\end{equation}
%

The usual scale employed to judge differences in
evidence from the models is the Jeffreys scale \cite{Jeffreys} which gives empirically calibrated levels of significance for the strength of evidence. 
In this work we will use a revisited and more conservative version of the Jeffreys convention suggested in \cite{Trotta} where 
$ \ln \mathit{B}_{ij}  = 0 - 1 $,
$ \ln \mathit{B}_{ij}  = 1 - 2.5 $,
$ \ln \mathit{B}_{ij}  = 2.5 - 5 $,
and $ \ln\mathit{B}_{ij}  > 5 $ indicate an {\textit{inconclusive}}, {\textit{weak}}, {\textit{moderate}} and {\textit{strong}} preference of the model $M_i$ with respect to the model $M_j$.
Note that, for an experiment which provides $\ln \mathit{B_{ij}} < 0$, it means support in favour of the reference model $M_j$  (see ref.~\cite{Trotta, Santos:2016sti} for a more complete discussion about this scale).
%
%
%
%
%
\begin{table}
\centering
\caption{{
$68\%$ confidence limits for the cosmological and collapse scheme parameters. 
The first columns-block refer to the minimal $\Lambda$CDM model; 
the second and third columns-block show the constraint on the {\it Newtonian} and {\it Wigner} scheme models; 
The $\Delta \chi^2_{best}$ and the $\ln {B}_{ij}$ refers to the difference with respect to the $\Lambda$CDM. }
\label{tab:Tabel_results}}
\scalebox{0.85}{
\begin{tabular}{c|cc|cc|cc}
\hline
\hline

\multicolumn{1}{c|}{$ $}&
\multicolumn{2}{c|}{\textbf{$\Lambda$CDM model}}& 
\multicolumn{2}{c|}{\textbf{Newtonian-scheme}}& 
\multicolumn{2}{c}{\textbf{Wigner-scheme}}
\\ 
Parameter	& mean & bestfit & mean & bestfit & mean & bestfit \\
\hline
$100\,\Omega_b h^2$ 	
& $2.222 \pm 0.022$ & $2.209 $ 		
& $2.222 \pm 0.020$ & $2.221 $
& $2.222 \pm 0.020$ & $2.218 $ 		
\\
$\Omega_{c} h^2$	
& $0.1197 \pm 0.0021$ & $0.1201 $	
& $0.1201 \pm 0.0020$ & $0.1201 $
& $0.1202 \pm 0.0019$ & $0.1194 $ 
\\
$100\, \theta$ 
& $1.04085 \pm 0.00045$ & $1.04114 $	
& $1.04083 \pm 0.00045$ & $1.04075 $
& $1.04081 \pm 0.00045$ & $1.04069 $
\\
$\tau$
& $0.077 \pm 0.018$& $0.070$	
& $0.086 \pm 0.018$& $0.087$
& $0.086 \pm 0.018$& $0.088$
\\
$n_s$ 
& $0.9654 \pm 0.0059$ & $0.9635 $	
& $0.9611 \pm 0.0056$ & $0.9598 $ 
& $0.9603 \pm 0.0058$ & $0.9648 $	
\\
$\ln 10^{10}A_s$  \footnotemark[1]
\footnotetext[1]{$k_0 = 0.05\,\Mpc^{-1}$.}
& $3.088 \pm 0.034$ & $ 3.080 $
& $4.130 \pm 0.021$ & $ 4.140$ 
& $2.850 \pm 0.022$ & $2.871$ 
\\
\A 
& $-$ & $-$ 									
& $-656.960 \pm 0.017$ & $-656.975 $				
& $-800.025 \pm 0.014$ & $-800.045 $				
\\
$100$ \B 	 
& $-$ & $-$
& $-0.0128\pm 0.0053$ & $-0.0107$ 
& $-0.011\pm 0.0044$ & $-0.0069$ 
\\
\hline
\hline
$\Delta \chi^2_{\rm best}$         
& & $-$	%
& & $3.4$
& & $1.4$
\\
$\ln \mathit{B}_{ij}$ \footnotemark[3]
\footnotetext[3]{The associated error is calculated with the simple error propagation formula, assuming that the two measurements are uncorrelated:  
$\sigma^2(\ln \mathit{B}_{ij})= \sigma^2(\ln \mathcal{E}_{i})+\sigma^2(\ln \mathcal{E}_{j})$}
& & $-$ 
& $-1.95 \pm 0.03$ &
& $-0.23 \pm 0.03$ &
\\
\hline
\end{tabular}}
\end{table} 
\begin{figure}
	\centering
	\includegraphics[width=0.7\hsize]{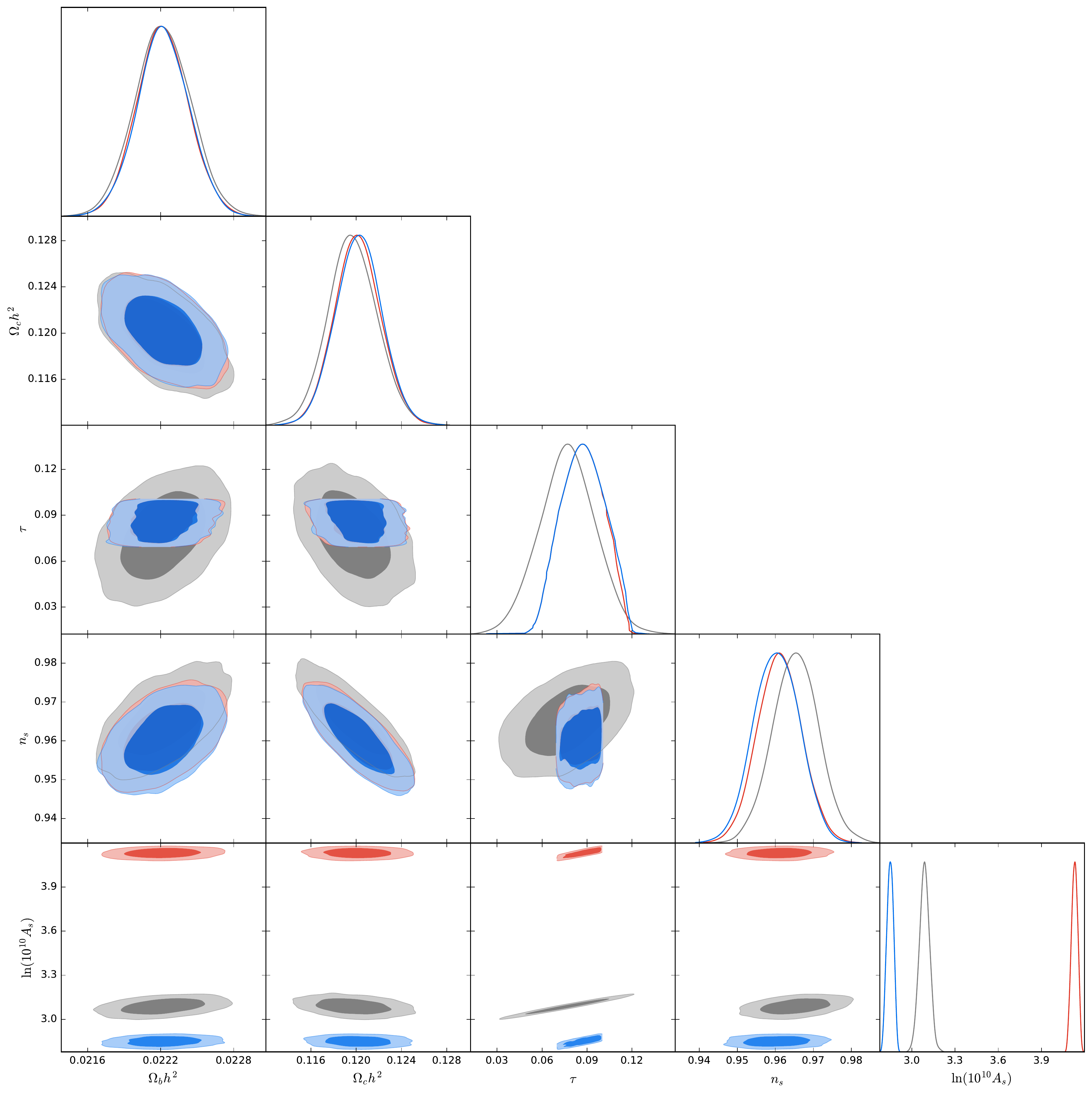}
	\caption{$68\%$ and $95\%$ confidence regions for the cosmological parameters of the $\Lambda$CDM (black line), {\it Newtonian} scheme (red line) and {\it Wigner} scheme (blue line) models using the Planck TT+lowP data. 
	The numerical results of these analyses are reported in Tab. \ref{tab:Tabel_results}.}
	\label{fig:Tiangle_plot}
\end{figure}
%
\section{Results}
\label{results}
The main quantitative results of our analysis are shown in Tab. \ref{tab:Tabel_results}, where we analysed the $\Lambda$CDM model, the Newtonian collapse and the {\it Wigner} collapse schemes.
We verify a significant agreement between the three models about the constraints on the cosmological parameters (see Fig.~\ref{fig:Tiangle_plot}).  For the collapse schemes, the $\tau$ parameter shows a preference for higher values while $n_s$ mean has a shift to lower values with respect to the $\Lambda$CDM model. The $\tau$ effect is due to the \A parameter behaviour and its degeneracies with the magnitude of the primordial spectra. 
Also the $A_s$ parameter assume different values, with a slight deterioration of the error at one sigma. 

\begin{figure}
	\centering
	\includegraphics[width=0.4\hsize]{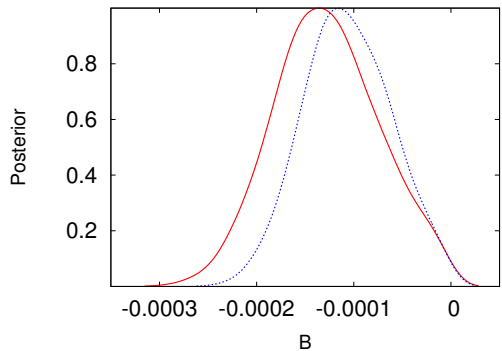}
	\caption{One-dimensional posterior probability densities for the \B parameter of the {\it Newtonian} scheme (red solid curve) and {\it Wigner} scheme (blue dotted).}
	\label{fig:Fig_posteriors}
\end{figure}
\begin{figure}
	\centering
	\includegraphics[width=0.4\hsize]{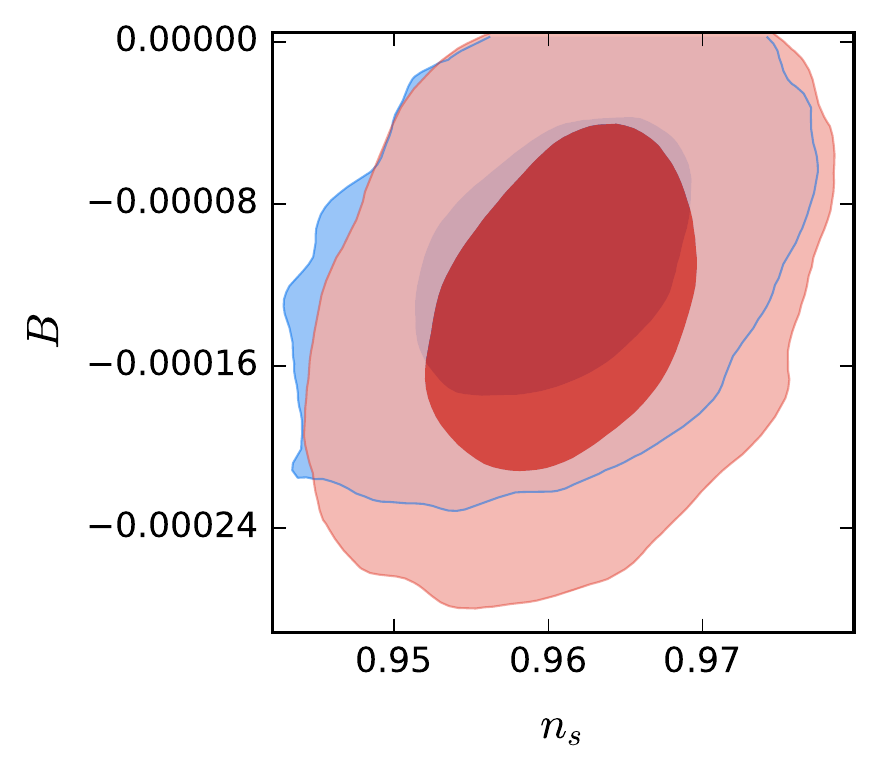}
	\caption{$68\%$ and $99\%$ two-dimensional confidence region in the $n_s - \mathcal{B}$ of the {\it Newtonian} scheme (red) and {\it Wigner} scheme (blue) analysis.}
	\label{fig:Fig_ns_B_2D}
\end{figure}
%
Remarkably, the data show a clear preference for non-zero values of \B, with the power-law form (recovered with $\mathcal{B}=0$) being excluded from the data within two sigma. 
It can be  clearly seen in Fig.(\ref{fig:Fig_posteriors}), where the Gaussian \B density posteriors distribution goes down for the zero value. The small improvement of the collapse scheme models on the $\chi^2$ values (with respect to the $\Lambda$CDM model) is displayed in the last rows of the Tab. \ref{tab:Tabel_results}.  In what concerns the Bayesian analysis, the CMB data give an inconclusive Bayesian evidence of the {\it Wigner} scheme with respect to the standard scenario, which amounts to saying that the $\Lambda$CDM model and the collapse {\it Wigner} scheme show the same Bayesian evidence. On the other hand, the {\it Newtonian} scheme model is weakly disfavoured with respect to the $\Lambda$CDM model, with $\mathit{\ln B}_{ij}= -1.95 \pm 0.03$. For completeness, we show in Fig.(\ref{fig:Fig_ns_B_2D}) the parameter space of $n_s$ and \B for both collapse schemes. We can see that the {\it Wigner} scheme constrains a smaller volume with respect to the {\it Newtonian} case, which may explain the difference of values of the Bayesian evidence, according to  Eq. (\ref{eq:bayes}).  We recall that the collapse models are penalized by the Bayesian analysis for having two extra parameters compared to the minimal standard $\Lambda$CDM model. However, due to the high non-linearity of the collapse parameter $\mathcal{A}$, results shown in Table \ref{tab:Tabel_results} should be regarded as the best fit for the {\it Newtonian} and {\it Wigner} schemes within the range $[-1000,0]$. Therefore, we can not reach any fair conclusion about the goodness of these models beyond this range of $\mathcal{A}$.   This can be clearly seen in Table~\ref{tab:Tabel_newtonian_other}, where we report the results obtained when two different values of $\mathcal{A}$ are considered for the {\it Newtonian} model. For completeness, we also report in Tables~\ref{tab:Tabel_results} and \ref{tab:Tabel_newtonian_other} the improvements in $\Delta \chi^2$ for the best fit collapse models with respect to the $\Lambda$CDM.

\begin{figure}
	\centering
	\includegraphics[width=0.9\hsize]{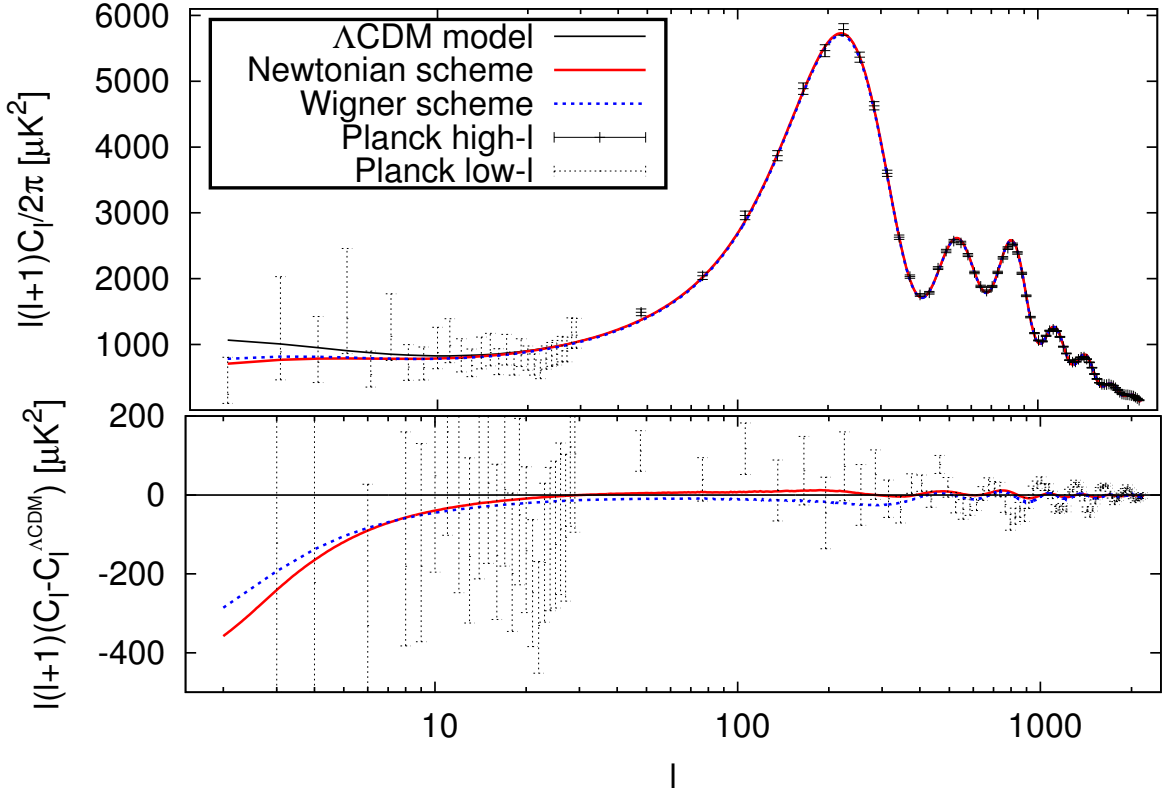}
	\caption{Anysotropy power spectrum for the $\Lambda$CDM (black line), the {\it Newtonian} (red solid line) and the {\it Wigner} scheme models (blue dotted line) for the best fit values reported in Tab.\ref{tab:Tabel_results}. In the bottom panel, the differential plot with respect to the $\Lambda$CDM best fit curve.}
	\label{fig:Fig_Anysotropy_TT}
\end{figure}

Finally, in Fig.(\ref{fig:Fig_Anysotropy_TT}) we show the anisotropy temperature power spectrum for the best fit values of the analysed models. Clearly, the collapse scheme scenarios are able to predict modifications in the low-$\ell$ region of the spectrum, which is not only in agreement with the data but also may be a possible explanation for the well-known lack of power at  high scale ($\ell<20$) \cite{Ade:2015xua, Copi:2013cya}.
%
%
\begin{table*}
\centering
\caption{
$68\%$ confidence limits the {\it Newtonian} scheme model fixing \A $=50$ and \A $=660$.
The $\Delta \chi^2_{best}$ and the $\ln {B}_{ij}$ refers to the difference with respect to the $\Lambda$CDM model and a negative value for the $\Delta \chi^2_{best}$ means a lower value for the $\Lambda$CDM model.}
\label{tab:Tabel_newtonian_other}
\begin{tabular}{c|cc|cc}
\hline
\hline

\multicolumn{1}{c|}{$ $}&
\multicolumn{2}{c|}{\textbf{\A = 50}}& 
\multicolumn{2}{c}{\textbf{\A = 660}}
\\ 
Parameter	& mean & bestfit & mean & bestfit \\
\hline
$100\,\Omega_b h^2$ 	

& $2.222 \pm 0.021$ & $2.222 $ 
& $2.222 \pm 0.021$ & $2.209 $ 
\\
$\Omega_{c} h^2$	

& $0.1192 \pm 0.0020$ & $0.1197 $
& $0.1201 \pm 0.0020$ & $0.1213 $ 
\\
$100\, \theta$ 

& $1.04091 \pm 0.00045$ & $1.04126 $
& $1.04084 \pm 0.00045$ & $1.04066 $
\\
$\tau$

& $0.084 \pm 0.018$& $0.072$
& $0.087 \pm 0.018$& $0.085$
\\
$n_s$ 

& $0.9697 \pm 0.0056$ & $0.9654 $ 
& $0.9616 \pm 0.0056$ & $0.9580 $	
\\
$\ln 10^{10}A_s$  \footnotemark[1]
\footnotetext[1]{$k_0 = 0.05\,\Mpc^{-1}$.}

& $4.073 \pm 0.0168$ & $ 4.058$ 
& $4.070 \pm 0.0170$ & $4.063$ 
\\

$100\,\ \mathcal{B}$ 	 

& $-0.0223\pm 0.0131$ & $-0.0003$ 
& $-0.0149\pm 0.0060$ & $-0.0195$ 
\\

\hline
\hline
$\Delta \chi^2_{\rm best}$         

& & $-0.6$ 
& & $3.4$ 
\\
$\ln \mathit{B}_{ij}$ \footnotemark[3]
\footnotetext[3]{The associated error is calculated with the simple error propagation formula, assuming that the two measurements are uncorrelated:  
$\sigma^2(\ln \mathit{B}_{ij})= \sigma^2(\ln \mathcal{E}_{i})+\sigma^2(\ln \mathcal{E}_{j})$}

& $-3.32 \pm 0.02$ & 
& $-1.89 \pm 0.02$ & 
\\
\hline
\end{tabular}
\end{table*} 
%
%
\section{Conclusions}

\label{conclusions}

Observations of the CMB radiation are one of the most powerful tools to study the physics of the early universe.  Starting with COBE's groundbreaking detection in the early nineties, the past two decades witnessed a great improvement in the measurements of the CMB fluctuations, which are now capable of ruling out theoretical models of inflation as well as some their alternatives. 

In this paper, we have studied the phenomenological predictions of the collapse models developed  in Ref. \cite{LLP15} considering only the case where the collapse happens before the horizon crossing. We have assumed a  more predictive parameterisation for the collapse time $\eta_c^k$, i.e.,  $\eta_c^k=\frac{A}{k}+\frac{B}{k^2}$, which is able to fit the observed lack of power at low multipoles of the CMB temperature auto-correlation angular power spectrum. We have performed a  statistical analysis to test  the observational viability of the so-called {\it Newtonian} and {\it Wigner} scheme scenarios in the light of the most recent CMB data, as recently reported by the Planck Collaboration. In order to compare the predictivity power of these models with respect to the  standard $\Lambda$CDM model, we have also performed a MCMC analysis and calculated the Bayesian evidence of each model.

The results, detailed in Sec. \ref{results}, show that collapse inflationary models can explain the current CMB data. Furthermore, we have obtained very restrictive bounds on the collapse parameter $\mathcal{B}$ and shown that its value is different from 0 at $2\sigma$ level. 
On the other hand, the values obtained for the usual cosmological parameters are consistent within $1 \sigma$ with those obtained by the Planck collaboration assuming a standard inflationary scenario. Finally, results from the Bayesian model comparison method show that the data can not distinguish between the $\Lambda${CDM} and the {\it Wigner} scheme collapse model, while the former model is preferred over the {\it Newtonian} scheme collapse model. 
%
\section*{Acknowledgments}
MB acknowledges financial support from the Funda\c{c}\~{a}o Carlos Chagas Filho de Amparo \`{a} Pesquisa do Estado do Rio de Janeiro (FAPERJ - fellowship {\textit{Nota 10}}). SL is  supported  by PIP 11220120100504  CONICET.  JSA is supported by Conselho Nacional de Desenvolvimento Cient\'{\i}fico e Tecnol\'ogico (CNPq) and FAPERJ. 
The authors acknowledge the use of Multinest code~\cite{Feroz:2008xx,Feroz:2007kg,Feroz:2013hea} and Gabriel Le\'on for useful discussions.

\bibliography{bibliografia}
\bibliographystyle{unsrtnat}

\end{document}